\newcommand{\ox}{\otimes}
\newcommand{\hc}{\dagger}
\newcommand{\ket}[1]{| #1 \rangle}
\newcommand{\bra}[1]{\langle #1 |}
\newcommand{\ktbr}[2]{\ket{#1}\bra{#2}}
\newcommand{\vxv}[1]{\ket{#1}\bra{#1}}

\newcommand\Tr{\mathop\mathrm{Tr}}
\newcommand\Id{\openone}
\newcommand{\tfrac}[2]{{\textstyle\frac{#1}{#2}}}

\newcommand\eq[1]{Eq.~(\ref{#1})}

\documentclass[showpacs,twocolumn,amssymb]{revtex4}
\begin{document}
\title{Affine-Detection Loophole in Quantum Data Processing}
\author{Alexander Yu.\ Vlasov}%
 \email{qubeat@mail.ru}
\affiliation{Federal Radiological Center (IRH),
197101 Mira Street 8, St.--Petersburg, Russia}
\date{11 Jan 2003}

\begin{abstract}
Here is considered a specific detection loophole, that is relevant not only
to testing of quantum nonlocality, but also to some other applications of
quantum computations and communications. It is described by a simple affine
relation between different quantum ``data structures'' like pure and mixed
state, separable and inseparable one. It is shown also, that due to such
relations imperfect device for a classical model may mimic measurements of
quantum correlations on ideal equipment.
\end{abstract}


\maketitle

\section{Introduction}
\label{sec:intro}

Let us consider a few standard formulae \cite{InvQC}. A density matrix for one
qubit can be represented as:
\begin{equation}
 \rho = \frac{1}{2}(\Id + a_1\sigma_x + a_2\sigma_y + a_3\sigma_z),
\end{equation}
where $\Id$ is the unit matrix, $\sigma_\mu$ are three Pauli matrices and
$\mathbf a =(a_1,a_2,a_3)$ is 3D vector of real parameters
$\|\mathbf a\| \le 1$. All pure states due to $\rho^2 = \rho$ satisfy to
simple condition
\begin{equation}
 \|\mathbf a\| = 1.
\label{a}
\end{equation}

It is clear, that any mixed state for such a simple model is related with
pure state via \emph{affine transformation}
\begin{equation}
 A_a:~~\rho' = a\rho + (1-a)\Id/N
\label{aff}
\end{equation}
where $a > 1$ is a real parameter, $N=2$ is dimension of Hilbert space for
example under consideration (cf \cite{SepVol,SepNMR,ClasNMR}). An inverse
transformation may be expressed using same formula with $1/a$,
\begin{equation}
A_a^{-1}=A_{1/a}
\label{affinv}
\end{equation}

For models with higher dimensional systems or more than one qubit the
transformation \eq{aff} maps to pure states only some class of mixed ones,
denoted as \emph{pseudo-pure} \cite{NMR}. The relation between pure and
pseudo-pure states discussed above devotes special consideration and
generalizations.

\section{Pseudo-pure states and ensemble computations}

Let us consider an unitary evolution (quantum gate) $U$:
\begin{equation}
 \rho \to U \rho U^\hc
\end{equation}
Such evolution does not change unit matrix $\Id$ and so we have a set
of similar models linked by the transformations $A_a$ \eq{aff}. Really the
models are equivalent in more rigor sense,
that can be described using a commutative diagram
\begin{equation}
 \begin{array}{ccc}
  \rho & \longrightarrow & U \rho U^\hc \\
 \Big\downarrow\rlap{$A$} & &\Big\downarrow\rlap{$A$}~ \\
  \rho' & \longrightarrow & U \rho' U^\hc
 \end{array}
\label{affdiagr}
\end{equation}
Here vertical arrows represent the affine transformation $A$ \eq{aff} and
commutativity means simply, that a path `right -- down' is equivalent to
`down -- right', i.e., \eq{aff} does not change a structure of the model.

In our case the affine equal models may be physically very different.
It is clear from an example with the Bloch sphere for one qubit. It is enough
to consider a sphere of pure states \eq{a} with radius one and thin layer with
radius $1/a$ representing some family of mixed states inside of this unit ball.
It is clear, that unitary transformations does not change this smaller sphere,
but it can be mapped to space of pure states by \eq{aff}.

Unlike of approach with vectors in Hilbert spaces, there only direction
does matter and length is fixed, here scale, ``length'' \eq{a} of traceless
part of density matrix is important parameter of the model.
Similar, but more difficult picture can be suggested for higher dimensional
and multi-qubit systems, where already not all mixed states could be used
for such a model. But for one qubit any mixed state is ``pseudo-pure.''

Let us consider now measurements in such a models. If to choose a measurement
basis with $N$ states $\ket{\mu_k}$, then probabilities are described as:
\begin{equation}
 p_k = \Tr\bigl(\rho\,\vxv{\mu_k}\bigr).
\end{equation}
The probabilities of corresponding outcomes for affine equal models are
related as
\begin{equation}
 p'_k = a p_k - (a-1)/N.
\label{paff}
\end{equation}

Due to \eq{paff} the measurements let us distinguish these models, but
similarity is such a strong, that in NMR quantum computation where $a$ is
\emph{very big} the mixed (pseudo-pure) states anyway
are used for simulations of quantum gates for pure states \cite{NMR}.
The NMR is one of the most developed branch of quantum
computations and it could be suggested, that in other areas such ``affine
loopholes'' also may have importance, but still did not investigated
with proper attention. It should be
mentioned yet, that problem discussed here is related also with quantum
communications and formally it is some kind of \emph{detection loophole}
(see below and also \cite{GisLoop1,GisLoop2}).

\section{General measurement problems}

The main problem here, that such loophole is rather hidden due to common
principles of any physical measurement. A measured value $X$ usually contains
two main kinds of systematic errors: rescaling $s$ and background noise $b$,
and so ``true'' value $X_0$ is calculated usually as
\begin{equation}
 X_0 = s\,(X - b).
\label{err}
\end{equation}
Both $X$ and $X_0$ here are classical values, there $X_0$ may be
considered as result of some imaginary ideal experiment, contrasting
with real raw outcome $X$.

The formula is appropriate for any measurements, but it is discussed here
in quantum mechanical framework, because it is similar with \eq{paff} and
overestimation of noise may produce specific problems, like treatment
mixed state as pure (discussed above) and separable state as inseparable.
Last case is discussed below and maybe even more important: it is known,
that specific quantum phenomena is related with nonlocality and inseparable
states. Local quantum system like qubit may be simulated by classical model
\footnote{It should be mentioned, that although the qubit can be modeled by
some classical program with a generator of random numbers for simulation of
a measurement, to treat it as ``a classical \emph{statistical} model'' is not
a very good idea. A difference is even more transparent in case of a qutrit
(a quantum system with three states) --- despite of possibility of a computer
simulation with a random number generator and few states (e.g., $2N^2-N$ for
an $N$-dimensional system \cite{Tens}) it may not be treated as a classical
statistical model due to the Kochen-Specker theorem. The resolution of this
apparent paradox is a fact, that not any formal manipulation with the random
numbers in an abstract and complicated algorithm corresponds to some
classical stochastic process (certainly, it may also does not correspond to
any quantum process, e.g., it can describe cloning, evolution of some other
universes or simply contains program errors).
}
and so testing quantum mechanics is related nowadays with different
kinds of nonlocal experiments. Overestimation of noise in \eq{err} can
produce a possibility to treat a classical process as a quantum one. It is
discussed below and clear already from well known notion, that classical
description of a quantum nonlocal process has a problem of \emph{``negative
probabilities''} \cite{FeySim}.

\section{Separability and Quantum Nonlocality}

It was described above, that the affine transformation \eq{aff} may maps any
state of qubit (or a pseudo-pure state of higher dimensional systems) to
pure one. But \eq{aff} may also describe transformations between other
classes of states.

In two-party quantum communications there is important notion of
\emph{separable} system \cite{InvQC} with density matrix represented as
\begin{equation}
 \rho = \sum_\mu \lambda_\mu \rho_\mu^{I} \ox \rho_\mu^{II},
\quad \lambda_\mu > 0.
\end{equation}

For a multi-party system the theory is more complicated due to absence of
an unique decomposition, but using the main idea, that a separable state can
be considered as a composition of \emph{classically correlated} systems,
it is possible to define it also as the similar sum with positive coefficients.
If it is not possible to represent density matrix in such a way, i.e.,
always appear some negative coefficients, the state is called
\emph{inseparable} (or entangled, but the first term seems more appropriate
in present context for mixed states).

It is reasonable in present paper to consider question about different
classes of states related by the affine transformation \eq{aff}
and an answer is rather trivial:
\begin{quote}
\emph{For arbitrary number of qubits any state can be produced from a
separable state by affine transformation \eq{aff} with some parameter $a$.}
\end{quote}
i.e., inseparable and even pure states can be produced from
``classically correlated states'' for some $a>1$
\cite{SepVol,SepNMR,ClasNMR}.

\noindent\textbullet~\textbf{Proof.} Let us consider
\emph{an inseparable state} with $n$ qubits. It is possible
to use the unique decomposition with up to $4^n$ tensor products of
$(\Id,\sigma_\mu)$ and rewrite it as some sum of up to $6^n$
tensor products of pure states
\begin{equation}
\rho_\mu^\pm \equiv \tfrac{1}{2}(\Id\pm\sigma_\mu).
\label{rhopm}
\end{equation}
It is enough to use expressions $\sigma_\mu=\rho^+_\mu-\rho^-_\mu$ and
$\Id = \rho^+_\mu+\rho^-_\mu$ (let it be $\Id = \rho^+_z+\rho^-_z$ for
certainty).
Let us consider now some negative term
\begin{equation}
-\alpha \rho_{\mu_1}^\pm\ox\cdots\ox\rho_{\mu_n}^\pm
\label{negterm}
\end{equation}
and rewrite it as
$$
\alpha (\Id-\rho_{\mu_1}^\pm\ox\cdots\ox\rho_{\mu_n}^\pm) - \alpha \Id,
$$
where $\Id\equiv\Id\ox\cdots\ox\Id$ can be represented as
\begin{equation}
\Id=
(\rho_{\mu_1}^++\rho_{\mu_1}^-)\ox\cdots\ox(\rho_{\mu_n}^+ +\rho_{\mu_n}^-).
\label{decomp}
\end{equation}
The \eq{decomp} with $2^n$ terms includes \eq{negterm} and
so instead of the negative term it is possible to write $2^n-1$ positive
terms and $-\alpha\Id$. Using such transformations it is possible to get rid
of all negative coefficients except of one for
$\Id\equiv\Id\ox\cdots\ox\Id$, but this last coefficient may be deleted using the
affine transformation \eq{aff}. It is enough to rewrite \eq{aff} as
$$
 \rho' = a(\rho+\frac{1-a}{N\,a}\Id),
$$
where $N=2^n$.
To delete negative term $-x\Id$, it is necessary to use the affine
transformation with
\begin{equation}
\frac{1-a}{N\,a} = x ~\Longrightarrow~ a=\frac{1}{1+N\,x}
\label{compensa}
\end{equation}
So the inseparable state can be transformed to a
separable one using \eq{aff} with $0<a<1$ and so an inverse
transformation \eq{affinv} with $a=N\,x+1$ is the necessary map from the
separable state.
\footnote{It is slightly refined version of proof suggested in
\cite{SepNMR}. Here is necessary to emphasize two properties of the
(over)complete set of projectors $\rho_\mu^\pm$ used in present
proof: (1) they linear span is whole space of $N \times N$ density
matrices and (2) any element of set belongs to some complete family
with $N$ orthogonal projectors. This proof may be applied to any
number of ``qu$N$its'' ($N$-dimensional systems) even with different
$N_k$, if to use higher-dimensional generalizations of the set $\rho_\mu^\pm$
with same properties (1, 2) \cite{Tens}. As an example may be used
\emph{``complete set of projectors''} with $2N_k^2-N_k$ elements for each
$N_k$-dimensional (sub)system described in \cite{Tens}. An ``existence''
proof of equivalent theorem for an arbitrary number of qu$N$its also can
be found in \cite{SepVol}.}
\textbullet

\smallskip

Let us consider an example with the Bell state
$\ket{\psi}=(\ket{01}-\ket{10})/\sqrt{2}$
\begin{widetext}
\begin{eqnarray*}
 \vxv{\psi} &=& \tfrac{1}{2}\bigl(\ktbr00\ox\ktbr11 + \ktbr11\ox\ktbr00
               -\ktbr10\ox\ktbr01 - \ktbr01\ox\ktbr10\bigr)\\
            &=& \tfrac{1}{2}(\rho_z^+\ox\rho_z^- + \rho_z^-\ox\rho_z^+)
-\tfrac{1}{4}(\sigma_x\ox\sigma_x+\sigma_y\ox\sigma_y)\\
           &=&\tfrac{1}{2}(\rho_z^+\ox\rho_z^- + \rho_z^-\ox\rho_z^+
  + \rho_x^+\ox\rho_x^- + \rho_x^-\ox\rho_x^+
   + \rho_y^+\ox\rho_y^- + \rho_y^-\ox\rho_y^+ - \Id),
\end{eqnarray*}
\end{widetext}
where $\rho_\mu^\pm$ are defined by \eq{rhopm} above.
Such decomposition is not unique and really it is always possible
to use no more than four terms. So six terms are more than necessary,
but it is appropriate for our purposes. Due to \eq{compensa} with $N=2^2=4$
and $x=1/2$ it is necessary to use $a=1/3$ to delete $-\Id/2$ and so the Bell
state may be produced using the affine transformation with $a=3$ from the
separable state
\begin{widetext}
\begin{equation}
\tfrac{1}{6}(\rho_z^+\ox\rho_z^- + \rho_z^-\ox\rho_z^+
+ \rho_x^+\ox\rho_x^- + \rho_x^-\ox\rho_x^+
+ \rho_y^+\ox\rho_y^- + \rho_y^-\ox\rho_y^+).
\label{qdice}
\end{equation}
\end{widetext}
It should be mentioned, that $a=3$ is not necessary a minimal possible value.
The parameter $a$ could be chosen arbitrary big using different mixed states.
So a measure of inseparability must be based on the \emph{lower} bound of $a$
between all possible separable states ({\em random robustness}
\cite{Vidal,DMN00,SK02}).

\section{Experiments, Tests, Classical Models}

Due to such relation some ``classically correlated states'' can be used
as a model of ``pure quantum'' system after appropriate distortion \eq{paff}
of measurement statistic.
Is it possible
to observe such pseudo-quantum effects due to errors of experimental
equipment or misinterpreting of data?
Formally, it could be really so, but here the model is more difficult than for
one system. Say, for measurement of intensity in one beam, the values \{$X_0$,
$X$, background $b$\} in \eq{err} may be simply associated with intensities
of different sources and the error discussed above may be related for example
with overestimation of the background intensity $b$.

It should be mentioned that responsible for such ``constant error'' $b$,
together with an unknown background noise and fixed leak of events per unit
time, could be also more difficult schemes.
Let us consider two complimentary events with probabilities $p_0$ and
$p_1 = 1-p_0$. A proportional lost of events $s$ is simply corrected due
to condition $p_0+p_1=1$ and as an example of other kind of errors can be
considered \emph{symmetric misclassification}
\begin{eqnarray*}
& p_i' = p_i + \tfrac{\varepsilon}{2}(p_i - p_{\,\bar{\imath}}) =
p_i+\tfrac{\varepsilon}{2}(2p_i-1) =
(1+\varepsilon)p_i-\tfrac{\varepsilon}{2},\\
& \bar{\imath} \equiv 1-i, \quad p_0'+p'_1=1.
\end{eqnarray*}
Such misclassification may be related for example with reduction of
sensitivity to one kind of events due to events of a complimentary kind,
but here it has a specific interest due to the precise correspondence
with \eq{paff} for $N=2$ and $a=1+\varepsilon$.

\smallskip

For a two-beam setup errors related with any separate beam may not
produce the discussed effect of ``pseudo-quantum nonlocality,'' but it can
appears in a count of a coincidence rate. Here is again the rescaling $s$
may not be relevant for such kind of error and it is related with
the background or threshold coefficient $b$ in \eq{err}.

Really any good experimentalist always feels such kind of problems even
without using of commutative diagrams like \eq{affdiagr} and produces
many test and calibrations of experimental equipment to get rid or estimate
$s$ and $b$ in \eq{err} together with more complicated sources of errors
\cite{Aspect}. But this affine detection loophole discussed here, because
it belongs to some general class of problems related with any observation
and experiment, when very different physical models may produce
mathematically almost equal behavior.

Let us compare it with well known Bell ideas \cite{BellSp} about necessity
of taking into account classical theories reproducing ``quantitative''
result of quantum mechanics for few fixed angles and with an absolute
different shape of a curve for the angular dependence (``saw'' instead
of ``cosine'', see \cite{BellSp} and also \cite{GisLoop1,GisLoop2}). It was
one of reason to use four different angles in such kind of experiments
\cite{Aspect,BellSp}. So it was interesting here to consider a classical
model with ``qualitative'' same behavior (e.g., it would be same ``cosine''
shape for the angular dependence, but with different absolute values,
cf \cite{GisLoop2}). Such apparent similarity may provokes wrong
interpretations.

\section{Conclusion}

In the paper was discussed problems related with specific
kind of \emph{affinity} \eq{aff} between different systems and models.

Say, using a ``broken'' device with a threshold:
\begin{equation}
 X = \left\{
\begin{array}{cl}
 X_0 - \theta;& X_0 \ge \theta\\
 0;        & X_0 < \theta
\end{array}
\right.
\end{equation}
and an ensemble of separable, i.e., ``classically correlated'' states,
we may produce same result as with an ideal measurement device for inseparable
states. For example state \eq{qdice} is separable and may be modeled by
classical local model with hidden variables, but due do specific
errors \eq{paff} the same system would mimic any statistical property of
inseparable Bell state, including Bell inequalities.

It is really disappointing fact, that an experimenter should use more
and more complicated equipment and measurement techniques to test quantum
correlations, that may be equivalent with a result of measurement of
classical correlations on a non-ideal device. It may be taken into account
in relation with questions about necessity of subtlest experiments. Maybe
it really still exists some place for doubts and loopholes.
Not in quantum mechanics, but in absolutely perfect tests using
the contemporary kind of \emph{classical} equipment.

\section*{Acknowledgements}
Author is grateful to Nicolas Gisin for some explanations,
to Andrey Grib, Valery Gorbachev, Alexey Lobashev, Alexander Chernitskii,
Ivan Sokolov, {\em et al} for useful comments and discussions on seminars
devoted to draft of present paper, and to Emanuel Knill and Martin Plenio
for information about important references.


\begin{thebibliography}{99}
\bibitem{InvQC} R. F. Werner,
\emph{E-print} \texttt{quant-ph/0101061} (2001).
\bibitem{SepVol}
K. \.{Z}yczkowski, P. Horodecki, A. Sanpera, and M. Lewenstein,
\emph{Phys. Rev. A} \textbf{58}, 883 (1998);
\emph{E-print} \texttt{quant-ph/9804024}.
\bibitem{SepNMR}
S. L. Braunstein, C. M. Caves, R. Jozsa, N. Linden, S. Popescu, and R. Schack,
\emph{Phys. Rev. Lett.} \textbf{83}, 1054 (1999);
\emph{E-print} \texttt{quant-ph/9811018}.
\bibitem{ClasNMR}
R. Schack and C. M. Caves,
\emph{Phys. Rev. A} \textbf{60}, 4354 (1999);
\emph{E-print} \texttt{quant-ph/9903101}.
\bibitem{NMR} D. G. Cory, R. Laflamme, E. Knill, \emph{et al},
\emph{E-print} \texttt{quant-ph/0004104} (2000).
\bibitem{GisLoop1}N. Gisin and B. Gisin,
\emph{Phys. Lett. A} \textbf{260}, 323 (1999);
\emph{E-print} \texttt{quant-ph/9905018}.
\bibitem{GisLoop2}N. Gisin and B. Gisin,
\emph{E-print} \texttt{quant-ph/0201077} (2002).
\bibitem{FeySim} R. P. Feynman,
\emph{Int. J. Theor. Phys.} \textbf{21}, 467 (1982).
\bibitem{Tens}  A. Yu. Vlasov,
\emph{E-print} \texttt{quant-ph/0104126} (2001).
\bibitem{Aspect} A. Aspect, J. Dalibard, and G. Roger,
\emph{Phys. Rev. Lett.} \textbf{49}, 1804 (1982).
\bibitem{BellSp}
J. S. Bell, \emph{Speakable and unspeakable in quantum mechanics},
(Cambridge University Press, Cambridge, 1987).
\bibitem{Vidal} G. Vidal and R. Tarrach,
\emph{Phys. Rev. A} \textbf{59}, 141 (1999);
\emph{E-print} \texttt{quant-ph/9806094} (1998).
\bibitem{DMN00} P. Deuar, W. J. Munro, and K. Nemoto,
{\em J. Opt. B: Quantum Semiclass. Opt.} {\bf 2}, 225 (2000);
\emph{E-print} \texttt{quant-ph/0002002} (2000).
\bibitem{SK02} C. Simon, J. Kempe,
\emph{Phys. Rev. A} \textbf{65}, 052327 (2002);
\emph{E-print} \texttt{quant-ph/0109102} (2001).
\end{thebibliography}
\end{document}